\documentclass[aps,prl,reprint,superscriptaddress,]{revtex4-1}
\usepackage{braket}
\usepackage{array}
\usepackage{times,amsmath,graphicx,amssymb}
\usepackage{color}
\usepackage[normalem]{ulem}
\usepackage{tikz}
\usepackage[pdfpagemode=UseNone,pdfstartview=FitH,colorlinks=true,linkcolor=blue,urlcolor=blue,anchorcolor=blue,citecolor=blue]{hyperref}
\usepackage{siunitx}
\usepackage{amssymb}
\definecolor{grey}{rgb}{.6,.6,.6}
\definecolor{forestgreen}{rgb}{.13,.55,.13}
\definecolor{amber}{rgb}{1.00,.31,.0}

\usetikzlibrary{positioning}

\begin{document}
\title{Unconventional Singularity in Anti-Parity-Time Symmetric Cavity Magnonics}

\author{Y. Yang}
\affiliation{Department of Physics and Astronomy, University of Manitoba, Winnipeg, Canada R3T 2N2}

\author{Yi-Pu Wang}
\email{Email: Yipu.Wang@umanitoba.ca}
\affiliation{Department of Physics and Astronomy, University of Manitoba, Winnipeg, Canada R3T 2N2}

\author{J.W. Rao}
\affiliation{Department of Physics and Astronomy, University of Manitoba, Winnipeg, Canada R3T 2N2}

\author{Y.S. Gui}
\affiliation{Department of Physics and Astronomy, University of Manitoba, Winnipeg, Canada R3T 2N2}

\author{B.M. Yao}
\affiliation{State Key Laboratory of Infrared Physics, Chinese Academy of Sciences, Shanghai 200083, People's Republic of China}

\author{W. Lu}
\affiliation{State Key Laboratory of Infrared Physics, Chinese Academy of Sciences, Shanghai 200083, People's Republic of China}

\author{C.-M. Hu}
\email{Email: hu@physics.umanitoba.ca}
\affiliation{Department of Physics and Astronomy, University of Manitoba, Winnipeg, Canada R3T 2N2}

\date{\today}
\begin{abstract}
	By engineering an anti-parity-time (anti-$\mathcal{PT}$) symmetric cavity magnonics system with precise eigenspace controllability, we observe two different singularities in the same system. One type of singularity, the exceptional point (EP), is produced by tuning the magnon damping. Between two EPs, the maximal coherent superposition of photon and magnon states is robustly sustained by the preserved anti-$\mathcal{PT}$ symmetry. The other type of singularity, arising from the dissipative coupling of two anti-resonances, is an unconventional bound state in the continuum (BIC). At the settings of BICs, the coupled system exhibits infinite discontinuities in the group delay. We find that both singularities co-exist at the equator of the Bloch sphere, which reveals a unique hybrid state that simultaneously exhibits the maximal coherent superposition and slow light capability.
\end{abstract}
\date{\today}
\maketitle

\textit{Introduction.---}
Singularities, which mark sudden changes that exhibit extraordinary behavior, are of broad interest. They often correspond to points where characteristic functions are not differentiable, or physical quantities approach infinity. They have great importance in both answering fundamental questions and realizing applications. For instance, the big-bang singularity in cosmology describes the origin of the universe \cite{PhysRevLett.96.141301,PhysRevD.65.103004,PhysRevD.8.4231,PhysRevLett.86.5227}. Van Hove singularities~\cite{PhysRev.89.1189} in crystals underly anomalies in frequency distribution functions~\cite{Li2010np,PhysRevB.81.155413,PhysRevLett.104.136803}. In photonics and wave physics, the most interesting singularities are exceptional points (EPs) and bound states in the continuum (BICs).

\begin{figure}[!t]
	\centering
	\includegraphics[width=0.42\textwidth]{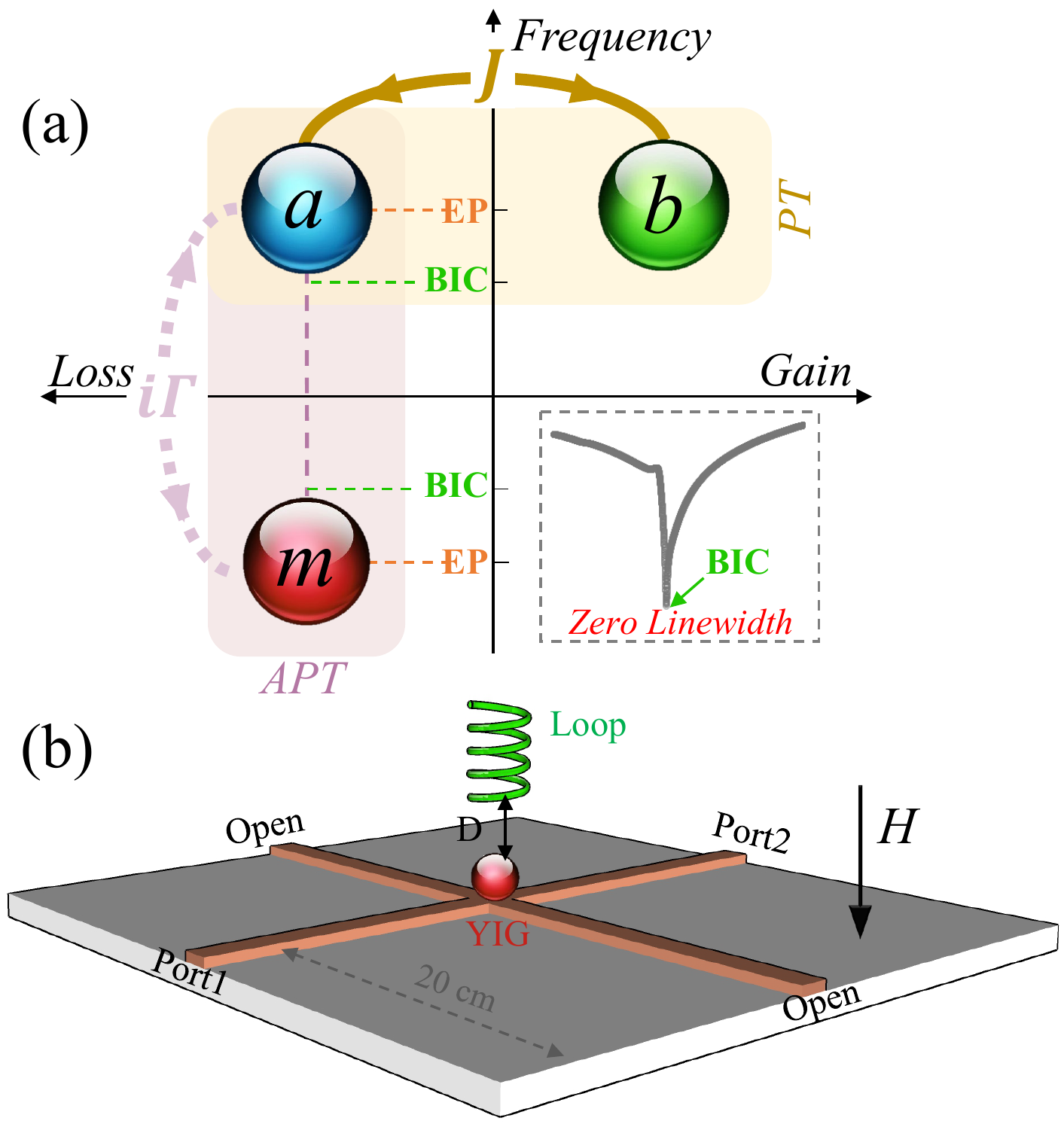}
	\caption{(a) In a $\mathcal{PT}$ symmetric system, two subsystems $a$ and $b$ share the same frequency and balanced loss and gain, and coherently coupled with each other at the rate $J$. In an anti-$\mathcal{PT}$ symmetric system, two subsystems $a$ and $m$ are frequency detuned but share the same loss (or gain), and they dissipatively coupled with each other at the rate $\Gamma$. The inset transmission amplitude spectrum indicates the profile of a BIC. (b) Schematic setup of the system designed to realize anti-$\mathcal{PT}$ symmetry and singularities. A YIG sphere is placed above the center of the crossline-microwave circuit to realize purely dissipative coupling between the cavity and magnon modes. A grounded loop antenna above the YIG sphere is used to control the damping rate of the magnon mode. External magnetic field $H$ is applied perpendicular to the cavity plane.}
	\label{Figure1}
\end{figure}

EPs are square-root singularities appearing in non-Hermitian systems. In general, such open systems have complex eigenvalues and non-orthogonal eigenvectors. However, it was found that real eigenvalues exist in parity-time ($\mathcal{PT}$) symmetric non-Hermitian systems~\cite{PhysRevLett.80.5243}, where EP emerges as the phase transition point between $\mathcal{PT}$ symmetry-preserved and $\mathcal{PT}$ symmetry-broken phases~\cite{RevModPhys.87.61,El-Ganainy2018,PhysRevX.9.041015}. It has led to many applications, such as topological states control \cite{Doppler2016,Xu2016,PhysRevX.8.031079} and highly sensitive detection \cite{Hodaei2017,Chen2017,PhysRevLett.123.213901}. In addition to $\mathcal{PT}$ symmetric systems, recent studies show that EPs also appear in anti-$\mathcal{PT}$ symmetric systems~\cite{ge2013antisymmetric,Peng2016,Choi2018,PhysRevLett.120.123902,Li170,PhysRevLett.124.030401,PhysRevLett.123.193604,PhysRevLett.124.053901,tserkovnyak2020exceptional,zhao2020observation}. Mathematically, an anti-$\mathcal{PT}$ symmetric Hamiltonian is obtained by multiplying the $\mathcal{PT}$ symmetric Hamiltonian by $i$, so that $\{\hat{H}^{(APT)},\mathcal{\hat{P}\hat{T}}\}=0$~\cite{supp}. As schematically shown in Fig.~\ref{Figure1}(a), to achieve a $\mathcal{PT}$ symmetric system, two subsystems with the same frequency must be coherently coupled (the coupling strength $J$ is real-valued) and feature balanced loss and gain. In contrast, realizing an anti-$\mathcal{PT}$ symmetric system requires two subsystems coupled by a purely dissipative mechanism (imaginary coupling strength $i\Gamma$) and featuring matched damping rates.

A BIC is a different kind of singularity. It was originally predicted by von Neumann and Wigner for electrons in an artificial complex potential~\cite{von1929}. Later, it was also found~\cite{Hsu2016} in photonics~\cite{PhysRevLett.100.183902,Kodigala2017,PhysRevLett.113.257401,Doeleman2018,liu2019circularly}, acoustics~\cite{PhysRevLett.118.166803,parker1966resonanc}, metamaterials and wave systems~\cite{PhysRevLett.118.267401,PhysRevApplied.12.014024,PhysRevLett.121.253901}. At BIC, a perfectly confined mode is embedded inside the radiation continuum, which can not radiate away. One prominent category of BICs, named Friedrich-Wintgen (FW) BICs, was previously proposed and understood as the result of destructive interference between two radiative modes~\cite{friedrich}. It has led to a wide range of applications, such as sensing and filtering~\cite{Hsu2016,PhysRevB.89.165111,Romano:19}, slow light~\cite{PhysRevLett.121.253901}, and quantum memory~\cite{bulgakov2015all}.

Despite broad interest in both EPs and BICs, creating a non-Hermitian system exhibiting both singularities is challenging, due to the fact that conventional coupled systems often lack the required eigenmodes tunability and controllability. In this letter, we engineer an anti-$\mathcal{PT}$ symmetric cavity magnonic system to solve this problem. Cavity magnonics systems are non-Hermitian that have attracted much attention recently~\cite{PhysRevLett.111.127003,PhysRevLett.113.083603,PhysRevLett.113.156401,PhysRevApplied.2.054002,PhysRevLett.114.227201,Zhangdk2017,cao2019exceptional,Lachance_Quirion_2019,zhang2019experimental,zhang2019quantum}. Built on magnon-photon couplings, such systems have great tunability. Serendipitously, dissipative coupling was recently discovered in cavity magnonics~\cite{PhysRevLett.121.137203}, which originates from the reservoir-mediated cooperative damping~\cite{PhysRevLett.123.127202,PhysRevX.5.021025,PhysRevB.101.064404,doi:10.1063/1.5144202}. These ingredients make cavity magnonics a promising platform for controlling EPs and BICs in the same system, as we demonstrate in this work. In addition to EPs that mark spontaneous anti-$\mathcal{PT}$ symmetry breaking, we discover an unconventional singularity in the anti-$\mathcal{PT}$ symmetry-preserved phase, which is a new kind of BIC that exhibits a unique combination of maximal coherent superposition and infinite discontinuities in the group delay.

\textit{Cavity Magnon System and Unconventional BICs                                                                                                                                  .---}
Cavity magnonics systems are described in general by the following Hamiltonian:
\begin{equation}\label{BICeq1}
\hat{H}_{0}/\hbar=\widetilde{\omega}_{c}\hat{a}^{\dagger}\hat{a}+\widetilde{\omega}_{m}\hat{m}^{\dagger}\hat{m}+(J+i\Gamma)(\hat{a}^{\dagger}\hat{m}+\hat{a}\hat{m}^{\dagger}),
\end{equation}
where $\widetilde{\omega}_{c}=\omega_{c}-i(\kappa+\beta)$ and $\widetilde{\omega}_{m}=\omega_{m}-i(\gamma+\alpha)$ are the uncoupled photon and magnon mode complex frequencies. $\hat{a}$~ ($\hat{a}^{\dagger}$) and $\hat{m}$~($\hat{m}^{\dagger}$) are the cavity photon and magnon annihilation (creation) operators. $\beta$ ($\kappa$) and $\alpha$ ($\gamma$) are the intrinsic (external) damping rates of the two modes. $J$ is the coherent magnon-photon coupling strength, and $i\Gamma\equiv i\sqrt{\kappa\gamma}$ is the dissipative coupling strength~\cite{PhysRevLett.123.127202}.

Setting $\beta=\alpha=0$, Eq.~(\ref{BICeq1}) results in the same form of the Friedrich-Wintgen Hamiltonian~\cite{Hsu2016,friedrich}. It sustains the FW-BIC at the condition of $ \omega_{m}-\omega_{c}=(\gamma-\kappa)J/\Gamma$, where one of the eigenvalues becomes purely real~\cite{Hsu2016}. Such a BIC has been realized in photonics~\cite{PhysRevLett.100.183902,PhysRevB.89.165111,Romano:19}, but it should not appear in cavity magnonics because $\alpha, \beta > 0$. However, we find that under the condition of $\kappa\gg(\alpha,\beta,\Gamma)\gg\gamma$, cavity magnonics systems are governed by the dissipative coupling of two anti-resonances~\cite{supp}, where a new type of BIC emerges as we explain below.

In the rotating frame with respect to the reference frequency $\omega_{ref}=(\omega_m+\omega_c)/2$, the purely dissipative coupling of anti-resonances is described by the following Hamiltonian in matrix form~\cite{supp}:
\begin{equation}\label{rot}
\hat{H}_{rot}/\hbar=\begin{bmatrix} -\frac{\Delta_H}{2}-i\beta & i\Gamma \\ i\Gamma &  \frac{\Delta_H}{2}-i\alpha \end{bmatrix},
\end{equation}
where $\Delta_H=\omega_m-\omega_c$ is the field detuning that is controlled by an external magnetic field $H$. The eigenvalues of the coupled system are found to be
\begin{equation}
\widetilde{\omega}_\pm=-i\frac{\alpha+\beta}{2}\pm\sqrt{(\frac{\Delta_H}{2}-i\frac{\alpha-\beta}{2})^2-\Gamma^2},
\label{eq4}
\end{equation}
with the corresponding eigenvectors given by
\begin{equation}
\ket{\lambda_\pm}
=\begin{pmatrix}
v_{1\pm}
\\
v_{2\pm}
\end{pmatrix}
=\begin{pmatrix}
i\Gamma
\\
\frac{\Delta_H}{2}-i\frac{\alpha-\beta}{2}\pm\sqrt{(\frac{\Delta_H}{2}-i\frac{\alpha-\beta}{2})^2-\Gamma^2}
\end{pmatrix}.
\label{eq5}
\end{equation}

Eq.~(\ref{eq4}) leads to two BICs appearing at the conditions of $\omega_{m}-\omega_{c} =\pm(\alpha+\beta)\sqrt{(\Gamma^2-\alpha\beta)/\alpha\beta}$, where one of the eigenvalues becomes purely real. Such cavity magnonic BICs are unconventional BICs. Intuitively, in systems with purely dissipative coupling ($J$ = 0), only one conventional FW-BIC would appear at $\omega_{m}-\omega_{c}$ = 0, where the dissipative coupling is an effective gain that cancels exactly the external damping of the coupled resonance~\cite{Hsu2016,friedrich}. In contrast, the unconventional BICs appear here as a pair, where the dissipative coupling cancels the intrinsic damping of the coupled anti-resonance.

The most interesting physics arises when the intrinsic damping rates are matched ($\alpha$ = $\beta$). In this case, EPs occur at $\Delta_H = \pm 2\Gamma$, where the two eigenvectors in Eq.~(\ref{eq5}) become parallel. Near the EPs, the cavity magnonic BICs appear at $\Delta_H = \pm 2\Gamma\sqrt{1-(\beta/\Gamma)^2}$, so that both singularities manifest in the same system. In such a case, the BICs emerge as singularities with unconventional character, which exhibit unique properties due to the preserved anti-$\mathcal{PT}$ symmetry, as we demonstrate below.

\textit{Experimental setup and methods.---}
The experimental setup is schematically shown in Fig.~\ref{Figure1}(b), a composite device consists of a crossline-microwave circuit, a 1-mm yttrium iron garnet (YIG) sphere, and a tunable loop antenna. The cavity mode, with the frequency $\omega_c/2\pi$ = 3.189 GHz and an intrinsic damping rate $\beta/2\pi$ = 10 MHz, is formed between the two open ports. The YIG sphere is glued 0.8 mm above the center of the crossline circuit. At this position, a purely dissipative magnon-photon coupling with $\Gamma/2\pi$ = 19.2 MHz is sustained by the microwaves traveling between Port 1 and Port 2 \cite{PhysRevLett.123.127202}. The external damping rates are found to be $\kappa/2\pi$ = 1 GHz and $\gamma/2\pi$ = 0.37 MHz for cavity and magnon modes, respectively. To realize the anti-$\mathcal{PT}$ symmetric cavity magnonics system, the small loop antenna is coupled to the YIG sphere to control its damping rate on demand. And the antenna is designed to have a negligible influence on the cavity mode. By changing the loop antenna's position, the total intrinsic damping rate of the magnon mode is tunable from $\alpha/2\pi$ = 2 MHz to 25 MHz, which enables the damping-matched condition to be realized at $\alpha/2\pi$ = $\beta/2\pi$ = 10 MHz~\cite{supp}.

\begin{figure}[t!]
	\centering
	\includegraphics[width=0.47\textwidth]{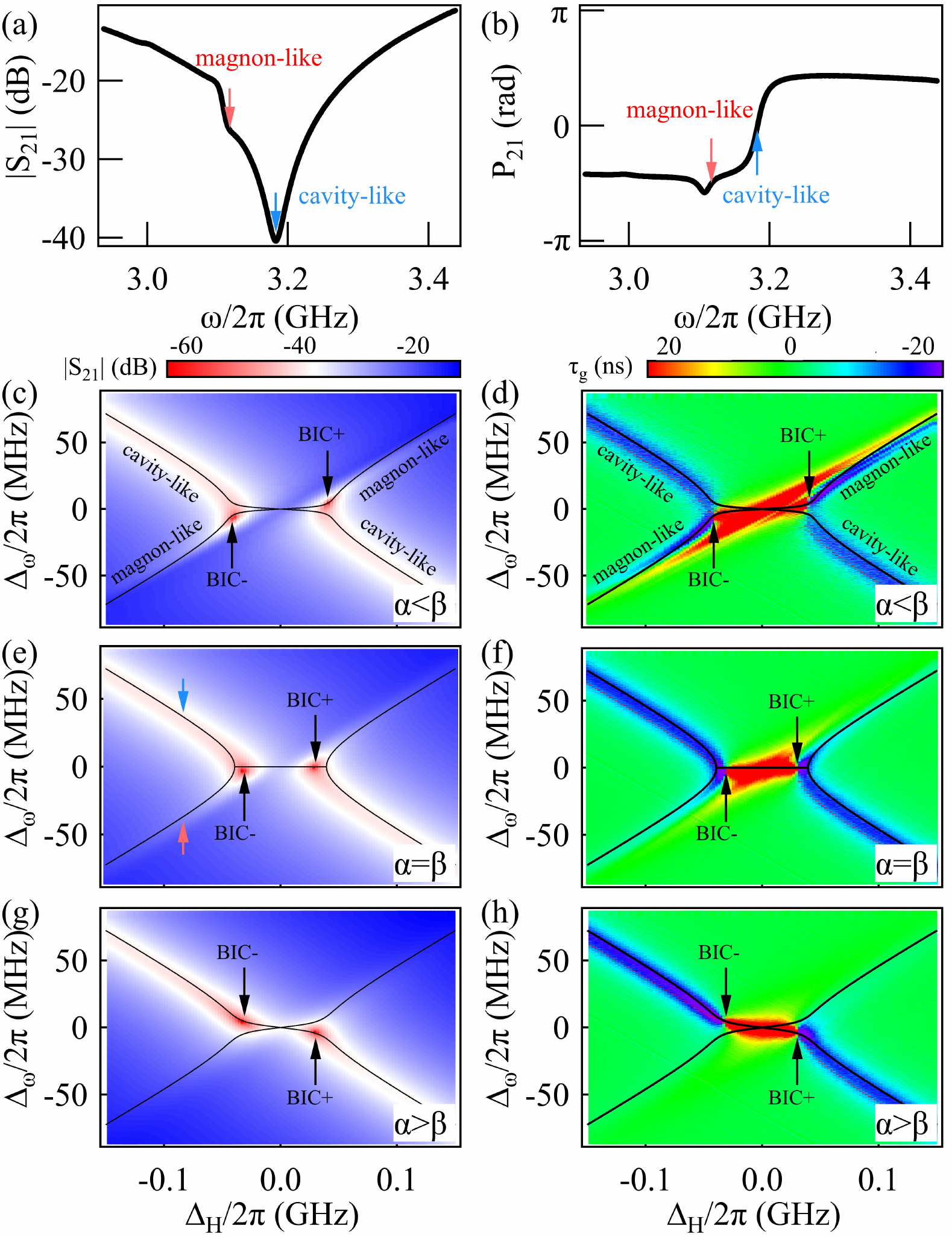}
	\caption{(a),(b) Typical transmission amplitude $|S_{21}|$ and phase $P_{21}$ at $\Delta_H/2\pi$=-80 MHz are plotted as a function of the probing frequency $\omega$. The red and blue arrows mark the magnon-like and cavity-like modes, respectively. (c), (e), and (g): The measured transmission amplitude $|S_{21}|$ mappings as a function of the frequency detuning $\Delta_{\omega}=\omega-\omega_{ref}$ and field detuning $\Delta_H$ in the conditions of $\alpha/2\pi$ = 5.6 MHz$<\beta/2\pi$, $\alpha/2\pi$ = 10 MHz$=\beta/2\pi$, and $\alpha/2\pi$ = 19.2 MHz$>\beta/2\pi$. The hot spots with the transmission amplitude approaches zero are two BICs. (d), (f), and (h): The group delay $\tau_g$ mappings. Two BICs emerge as singularities, where the group delay abruptly switches from negative infinity (purple) to positive infinity (red). The vertical red and blue arrows indicate the detuning setting for (a) and (b). The calculated eigenfrequency dispersions are co-plotted as black curves.}
	\label{Figure2}
\end{figure}

Transmission measurements are performed by connecting Port 1 and Port 2 to a vector network analyzer (VNA). The external magnetic field $H$ is applied perpendicularly to the circuit plane, which saturates the magnetization of the YIG sphere and sets the magnon mode frequency $\omega_m$. By sweeping the magnetic field and by changing the position of the loop antenna, this setup enables us to precisely control the eigenmodes in the two-dimensional parameter space defined by field detuning $\Delta_H$ and damping difference $\alpha-\beta$.

%\textit{Anti--$\mathcal{PT}$ symmetry and methods.---}

\textit{Experimental results.---}
Firstly, we display data that unveils the singularity nature of the cavity magnonic BIC. Transmission spectra $S_{21}(\omega) =|S_{21}|e^{-iP_{21}}$ are measured as a function of the frequency detuning $\Delta_{\omega}=\omega-\omega_{ref}$ and field detuning $\Delta_H$. Three typical damping conditions are selected by adjusting the position of the loop antenna, with $\alpha < \beta$,  $\alpha = \beta$, and $\alpha > \beta$. Typical amplitude $|S_{21}|$ and phase $P_{21}$ measured at $\Delta_H/2\pi$ = -80 MHz in condition $\alpha = \beta$ are plotted in Figs.~\ref{Figure2}(a) and \ref{Figure2}(b), respectively. Figs.~\ref{Figure2}(c), 2(e), and 2(g) show the mappings of $|S_{21}|$. The calculated eigenfrequencies are co-plotted as solid curves. The dissipatively coupled system clearly displays level attraction.  Two BICs appear as hot spots where the amplitude dip approaches zero, and the unloaded $Q$-factor approaches infinity~\cite{supp}. A more intriguing feature is revealed from the mappings of the group delay $\tau_g = -\partial P_{21}/\partial \omega$ that are plotted in Figs.~\ref{Figure2}(d), 2(f), and 2(h). Here, BICs emerge as singularities where the group delay of the coupled system abruptly switches from negative infinity to positive infinity~\cite{supp}. At the BICs, the group velocity $v_g \propto 1/\tau_g$ is zero, revealing that the cavity magnonic BIC is a singularity that can be utilized for slowing microwave photons.

\begin{figure}[t!]
	\centering
	\includegraphics[width=0.47\textwidth]{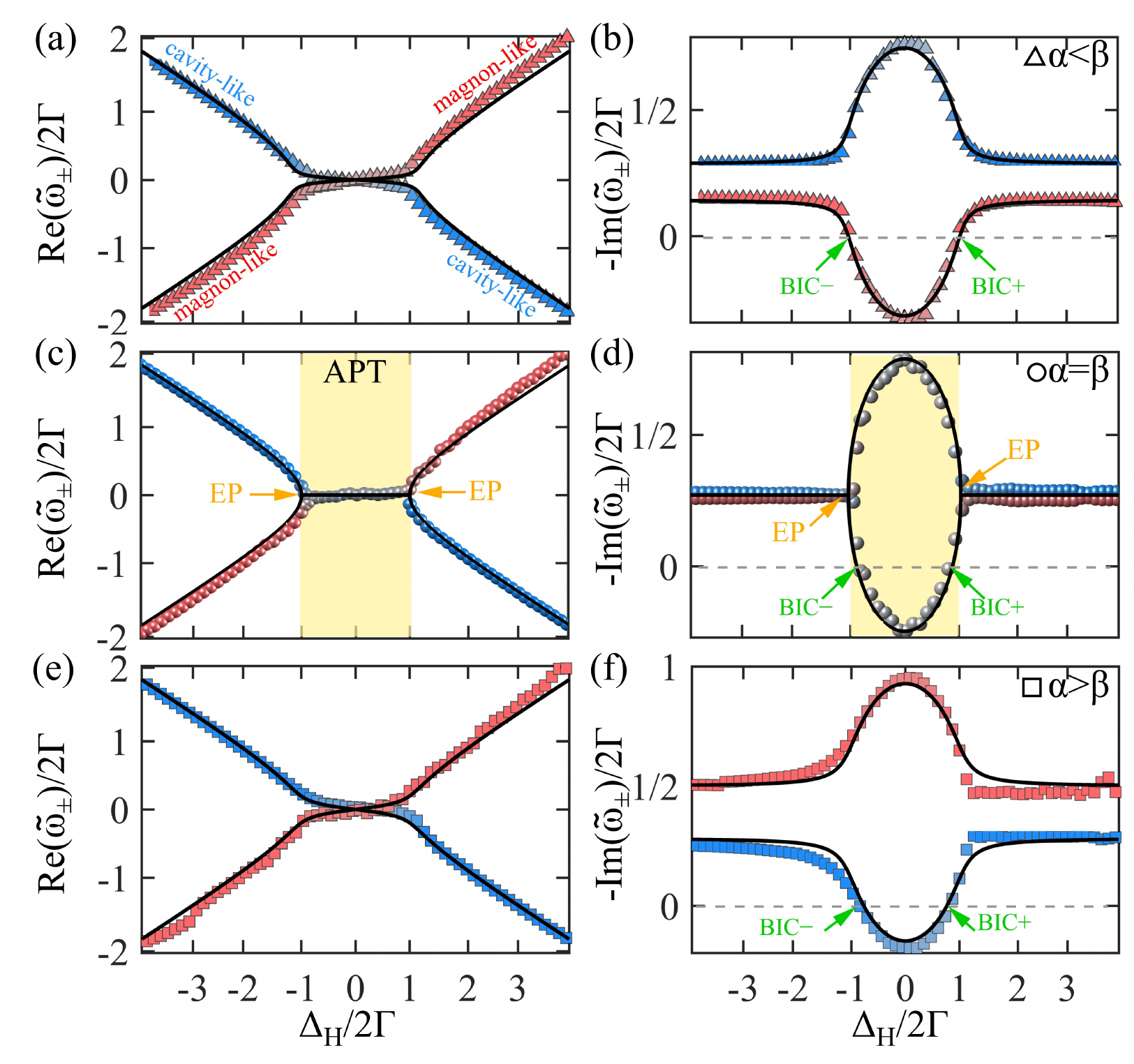}
	\caption{(a), (c), and (e): The real parts of the eigenvalues as a function of the field detuning $\Delta_{H}$ in the conditions of $\alpha<\beta$ (triangles), $\alpha=\beta$ (circles), and $\alpha>\beta$ (squares). (b), (d), and (f): The imaginary parts of the eigenvalues in corresponding conditions. The eigenvalues are colored by the dominance of cavity mode (blue) and magnon mode (red), grey color is the 50:50 dominance in between. The BICs are observed at zero damping and marked by green arrows. In condition $\alpha = \beta$, for $|\Delta_H/2\Gamma|<1$, the system is in the anti-$\mathcal{PT}$ symmetry-preserved phase (shaded in gold and marked as APT). For $|\Delta_H/2\Gamma|>1$, the system is in the anti-$\mathcal{PT}$ symmetry-broken phase. The EPs are observed at $|\Delta_H/2\Gamma|=1$ and marked by orange arrows. The calculated eigenfrequency dispersions are co-plotted as black curves.}
	\label{Figure3}
\end{figure}

Next, we demonstrate the simultaneous realization of both EPs and BICs by analyzing the fitted eigenvalues.  Real and imaginary parts of eigenvalues are plotted as a function of the field detuning, as shown in Fig.~\ref{Figure3}. The eigenvalues calculated from Eq.~(\ref{eq4}) are co-plotted as solid lines, which agree well with the measured data. The most striking features are seen at the damping-matched condition $\alpha = \beta$, as shown in Figs.~\ref{Figure3}(c) and \ref{Figure3}(d). Here, interlaced degeneracy and divergency appear. For $|\Delta_H/2\Gamma|<1$, the real parts of the eigenvalues are degenerate and equal to zero. It means that in the lab frame, both eigenfrequencies are locked to the reference frequency $\omega_{ref}$. The imaginary parts are divergent. Hence, the parameter regime of $|\Delta_H/2\Gamma|<1$ at $\alpha$ = $\beta$ is the anti-$\mathcal{PT}$ symmetry-preserved phase (marked as APT), where the eigenvalues are purely imaginary. Beyond this regime for $|\Delta_H/2\Gamma|>1$, the real parts are divergent while the imaginary parts are degenerate. These are the anti-$\mathcal{PT}$ symmetry-broken phases, where the eigenvalues are complex. Two EPs appear at $\Delta_H = \pm2\Gamma = \pm$ 38.4 MHz, and spontaneous symmetry breaking takes place. Furthermore, two BICs are observed at $\Delta_H = \pm 2\Gamma\sqrt{1-(\beta/\Gamma)^2} \approx \pm$ 32.0 MHz, where Im$(\widetilde{\omega}_{\pm})=0$ as depicted in Fig.~\ref{Figure3}(d).

When the damping-matched condition is violated as $\alpha\neq\beta$, the anti-$\mathcal{PT}$ symmetry-preserved phase diminishes, so that EPs that were observed in Fig.~\ref{Figure3}(c) merge at $\Delta_H$ = 0 in Figs. \ref{Figure3}(a) and \ref{Figure3}(e). Here, the eigenvalues remain purely imaginary, but the two eigenvectors no longer coalesce. This means the EPs disappear in damping-mismatched conditions. In contrast, we find that BICs, whose origin is conceptually different from EPs, are preserved at the lower damping branch, as shown in Figs.~\ref{Figure3}(b) and \ref{Figure3}(f). However, we note that no BIC shows up in $\mathcal{PT}$ symmetric systems \cite{PhysRevLett.80.5243,RevModPhys.87.61,El-Ganainy2018,PhysRevX.9.041015,Doppler2016,Xu2016,PhysRevX.8.031079,Hodaei2017,Chen2017,PhysRevLett.123.213901}.

Now, by utilizing the excellent eigenspace controllability of our platform, we construct the Riemann surface and Bloch sphere, to reveal the unique properties of the unconventional BICs in the anti-$\mathcal{PT}$ symmetry-preserved phase.

\begin{figure}[t!]
	\centering
	\includegraphics[width=0.45\textwidth]{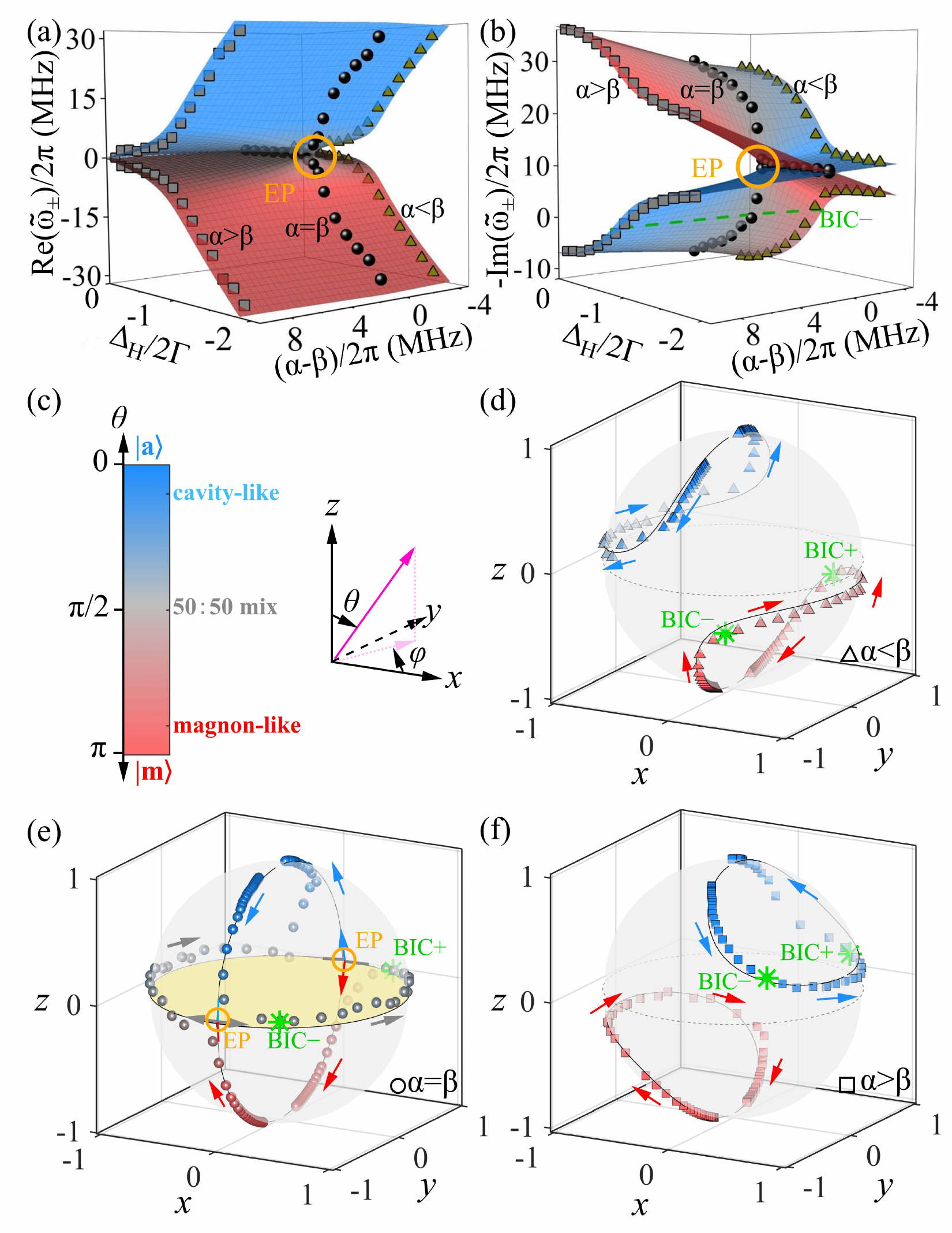}
	\caption{The Riemann surfaces of the calculated (a) real and (b) imaginary eigenvalues in the parameter space of field detuning $\Delta_H$ and damping difference $\alpha-\beta$. The EPs are labeled by orange circles, and the measured eigenvalues of conditions $\alpha < \beta$, $\alpha = \beta$, and $\alpha > \beta$ are added as triangles, circles, and squares. The position of BIC- is emphasized as green dashed line. Only negative field detuning data are displayed for clarity. (c) We construct the Bloch spheres in the spherical coordinates $\theta,\varphi$ that related to Cartesian coordinates $x,y,z$. The polar and azimuthal angles are defined in the text. The colorbar is defined by the relative intensity $\theta$, which shows the dominance of cavity mode $\ket{a}$ (blue) and magnon mode $\ket{m}$ (red), with the grey color as the 50:50 dominance in between. (d), (e), and (f): The Bloch spheres of conditions $\alpha < \beta$, $\alpha = \beta$, and $\alpha > \beta$. The colored arrows indicate the direction of increasing $\Delta_H$. The EPs and BICs are labeled by the orange circles and the green stars. The equator of the Bloch sphere with $\theta = \pi/2$ is marked by the grey dashed lines. The anti-$\mathcal{PT}$ symmetry-preserved phase on the equator is shaded in gold in condition $\alpha = \beta$.}
	\label{Figure4}
\end{figure}

The Riemann surfaces of the calculated real and imaginary eigenvalues are plotted in Figs.~\ref{Figure4}(a) and \ref{Figure4}(b), respectively. For clarity, we only plot the negative field detuning part ($\Delta_H < 0$). The EPs are displayed as the merging points on the surfaces and labeled by open circles. The locations of BICs are indicated by the dashed line. The measured eigenvalues in the conditions of $\alpha < \beta$, $\alpha = \beta$, and $\alpha > \beta$ are plotted with markers. Clearly, by changing the field detuning and damping difference, we are able to set the eigenvalues at anywhere on the Riemann surfaces, so that the system can be easily used for topological modes control by encircling the EPs on the Riemann surfaces.

The Bloch sphere is constructed from the eigenvectors $\ket{\lambda_{\pm}}= (v_{1\pm}, v_{2\pm})^{T}$. By using the complex notation $v_{1}=|v_{1}|\cdot e^{i\varphi_{1}}$ and $v_{2}=|v_{2}|\cdot e^{i\varphi_{2}}$, we define relative intensity $\theta =2\arctan(|v_{2}|/|v_{1}|)$ and relative phase $\varphi$ = $\varphi_{1}-\varphi_{2}$ as the polar and azimuthal angles, respectively. Mapping each eigenstate to a set of $(\theta,\varphi)$, the Bloch sphere is constructed where the uncoupled cavity mode $\ket{a}$ and magnon mode $\ket{m}$ are the north and south poles, respectively. The measured eigenvector evolution in the conditions of $\alpha < \beta$, $\alpha = \beta$, and $\alpha > \beta$ are plotted in Figs.~\ref{Figure4}(d), \ref{Figure4}(e), and \ref{Figure4}(f) on the Bloch sphere, where the arrows indicate the direction of increasing the field detuning $\Delta_H$.

Again, the most striking features are seen at the condition $\alpha = \beta$ as shown in Fig.~\ref{Figure4}(e). Here, with increasing $\Delta_H$, the cavity-like mode evolves downward from the north pole, while the magnon-like mode evolves upward from the south pole. At $\Delta_H=-2\Gamma$, two modes coalesce at the first EP on the equator. Further increasing $\Delta_H$, they separate and move along the equator in opposite directions. On the equator, one of the hybridized modes turns into BICs at $\Delta_H = \pm 2\Gamma\sqrt{1-(\beta/\Gamma)^2}$, before it rejoins the other mode at $\Delta_H=2\Gamma$ where the second EP is located. Afterwards, both modes return to their original poles. The result shows that in the anti-$\mathcal{PT}$ symmetry-preserved phase, both hybridized modes are \textquotedblleft equator modes" that share 50-50 contributions from cavity photon and magnon modes. Such equator modes are non-trivial. They stem from the maximal coherent superposition of photons and magnons, and they are robustly sustained over a broad detuning range in the anti-$\mathcal{PT}$ symmetry preserved phase. Hence, our result reveals a unique hybridized state that combines the maximal coherent superposition and zero group velocity. Such an unconventional singularity may find great applications in quantum information processing. One of the immediate capabilities is slowing light while preserving maximal magnon-photon coherence.

When $\alpha\neq\beta$, the evolution traces of the cavity-like and magnon-like modes are two closed circles, separated from each other in two hemispheres.  In these conditions, none of the eigenvectors reach the equator, so that EPs never emerge. In contrast, BICs still appear on the hemisphere for the lower damping mode branch. The co-existence of both kinds of singularities only takes place in the damping-matched system, where both EPs and BICs are constituted by the equator modes.

\textit{Conclusion.---}An anti-$\mathcal{PT}$ symmetric cavity magnonics system is engineered for the study of non-Hermitian singularities. Controlled evolutions of eigenmodes are demonstrated on both the Riemann surface and Bloch sphere. Two conceptually different singularities, namely EPs and BICs, are both observed on the equator of the Bloch sphere in the damping matched condition. Our work shows that although $\mathcal{PT}$ and anti-$\mathcal{PT}$ symmetric systems have similar eigen-structures, the manifestation of singularities in the two systems are remarkably different, which is a signature of distinct dynamics. We anticipate that, due to the co-existence of different types of singularities, anti-$\mathcal{PT}$ symmetric cavity magnonic systems could be even more useful than versatile $\mathcal{PT}$ symmetric systems. This may inspire cross applications that combine diverse capabilities such as eigenmode control, coherent state preparation, singularity engineering, and microwave confinement.

\begin{acknowledgments}

This work has been funded by NSERC Discovery Grants and NSERC Discovery Accelerator Supplements (C.-M. H.). We thank Y.T. Zhao and J. Burgess for discussions and helps.

\end{acknowledgments}

\end{document}